\documentclass[aps, showpacs, showkeys,nofootinbib,floatfix]{revtex4}

\usepackage{amssymb}
\usepackage{amsmath}
\usepackage{graphicx}
\usepackage[dvipdfm,colorlinks,citecolor=blue,urlcolor=blue]{hyperref}

\begin{document}
\title{\textbf{Conditions and instability in $f(R)$ gravity
with non-minimal coupling between matter and
geometry}\footnote{\small{Supported by the National Natural Science
Foundation of China (Grant No.10875056,$10932002$ and $10872084$),
and the Scientific Research Foundation of the Higher Education
Institute of Liaoning Province, China (Grant No. 2009R35 and
2007T087).}}}

\author{Jun Wang$^1$ \footnote{E-mail: wangjun\_3@126.com}, Ya-Bo Wu$^1$
 \footnote{Corresponding author: ybwu61@163.com}, Yong-Xin Guo$^2$, Fang Qi$^1$, Yue-Yue Zhao$^1$, Xiao-Yu Sun$^1$}

\affiliation{$^1$Department of Physics, Liaoning Normal
University, Dalian 116029, P.R.China, $^2$College of Physics, Liaoning University, Shenyang 110036,
P.R.China}

\begin{abstract}
In this paper on the basis of the generalized $f(R)$ gravity model
with arbitrary coupling between geometry and matter, four classes of
$f(R)$ gravity models with non-minimal coupling between geometry and
matter will be studied. By means of conditions of power-law
expansion and the equation of state of matter less than $-1/3$, the
relationship among $p$, $w$ and $n$, the conditions and the
candidate for late-time cosmic accelerated expansion will be
discussed in the four classes of $f(R)$ gravity models with
non-minimal coupling. Furthermore, in order to keep considering
models to be realistic ones, the Dolgov每Kawasaki instability will
be investigated in each of them.
\end{abstract}


\maketitle

\section{$\text{Introduction}$}
~~~~According to recent observational data sets\cite{1,2,3}, our
current universe is flat and undergoing a phase of the accelerated
expansion which started about five billion years ago. To explain
this phenomena, a variety of models have been proposed which may be
divided into three broad classes. First, it is possible that there
is some undiscovered property in our existing model of gravity and
matter that leads to acceleration at the present time. In thses
scenarios, one might consider including the existence of a tiny
cosmological constant and the possibility of the backreaction of
cosmological perturbations.

Second is the idea that the Universe is dominated by an exotic
component with large negative pressure, usually referred to as dark
energy. The simplest form of dark energy is cosmological constant
$\Lambda$ which would encounter ※fine-tuning§ problem and
※coincidence§ problem. Other valid dark energy models are provided
by scalar fields, such as: Quintessence\cite{4,5}, which is
introduced to solve the ※coincidence§ problem and characterized by
the equation of state (EOS) $w_{de}$ between -1 and -1/3 (namely,
$-1<w_{de}<-1/3$); Phantom (ghost) field\cite{6}, which owns a
negative kinetic energy and characterized by the EOS $w_{de}$ less
than -1 (namely, $w_{de}<-1$); Tachyon field\cite{7,8}which can act
as a source of dark energy depending upon the form of the tachyon
potential, and so on. Other scenarios on dark energy include brane
world\cite{9}, generalized Chaplygin gas\cite{10}, holographic dark
energy\cite{11}, etc. Unfortunately, up to now a satisfactory answer
to the question that what dark energy is and where it came from has
not yet to be obtained.

Finally, eliminating the need of dark energy, one may consider
modified theories of gravity in which the late-time cosmic
accelerated expansion can be realized by an infrared modification.
There are numerous ways to deviate from Einstein's theory of
gravity. The most well-known alternative to General Relativity (GR)
is scalar-tensor theory\cite{12,13}. There are still numerous
proposals for modified theories of gravity in contemporary
literature, such as DGP (Dvali- Gabadadze-Porrati) gravity\cite{14},
braneworld gravity\cite{15}, TeVeS (Tensor-Vector- Scalar)\cite{16},
$f(R)$ theories of gravity\cite{17}, Einstein-Aether theory\cite{18}
and so on.

Among these theories, $f(R)$ gravity is very competitive. Here
$f(R)$ is an arbitrary function of the Ricci scalar $R$. One can add
any form of $R$ in it, such as $1/R$\cite{19} (the simplest one),
$\ln R$\cite{20}, positive and negative powers of $R$\cite{21},
Gauss-Bonnet invariant\cite{22}, etc. The more general forms of
$f(R)$ can be considered including coupling between $f(R)$ and
scalar\cite{23}, multidimensional $1/R$ theory\cite{24} and so on.
In $f(R)$ theories of gravity, the expansion history of the universe
is naturally explained by the fact that some gravitational terms
which support the inflation at early-time universe, while other
terms which cause the cosmic acceleration at late-time universe. It
is worth stressing that considering some additional conditions, the
early-time inflation and late-time acceleration can be unified by
different role of gravitational terms relevant at small and at large
curvature. However, $f(R)$ gravity is not perfect because of
containing a number of instabilities. For instance, the theory with
$1/R$ may develop the instability\cite{25}. But by adding a term of
$R^{2}$ to this specific $f(R)$ model, one can remove this
instability\cite{20,21}. For more general forms of $f(R)$, the
stability condition $f'' \geq 0$ can be used to test $f(R)$ gravity
models\cite{26}.

Recently, a general model of $f(R)$ gravity has been proposed in
Ref.\cite{27}, which contains a non-minimal coupling between
geometry and matter. This coupling term can be considered as a
gravitational source to explain the current acceleration of the
universe. The viability criteria for such a theory was recently
discussed in Refs.\cite{27,28}. However, a more general model, in
which the coupling style is arbitrary and the Lagrangian density of
matter only appears in coupling term, has been proposed in
Ref.\cite{29} and it can represent the former case. The purpose of
this paper is to discuss the conditions for late-time behaviour in
$f(R)$ gravity with non-minimal coupling between matter and
geometry.

This paper is organized as follows. In the next section, for the
general $f(R)$ gravity models with arbitrary and non-minimal
couplings, the field equations, the energy conditions and the
Dolgov每Kawasaki instability will be given, respectively. In this
class of models, the energy-momentum tensor of matter is generally
not conserved due to the appearance of an extra force as mentioned
in Ref.\cite{27}. The conditions for late-time cosmic accelerated
expansion and the instability in $f(R)$ gravity with non-minimal
coupling will be discussed in sections $3$ and $4$, respectively.
Four classes of models will be taken into consideration in those two
sections. Summary is given in the last section.

\section{The general $f(R)$ gravity with coupling between matter and geometry}

~~~~A more general action in $f(R)$ gravity, in which the coupling
style between matter and geometry is arbitrary and the Lagrangian
density of matter only appears in coupling term, is given by
\begin{equation}\label{1}
S=\int[\frac{1}{2}f_{1}(R)+G(L_{m})f_{2}(R)]\sqrt{-g}d^{4}x,
\end{equation}
where we have chosen $\kappa = 8\pi G= c = 1$, which we shall adopt
hereafter. $f_{i}(R)$ ($i = 1, 2$) and $G(L_{m})$ are arbitrary
functions of the Ricci scalar $R$ and the Lagrangian density of
matter respectively. When $f_{2}(R) = 1$ and $G(L_{m}) = L_{m}$, we
obtain the general form of $f(R)$ gravity with non-coupling between
matter and geometry. Furthermore, by setting $f_{1}(R) = R$, action
(\ref{1}) can be reduced to the standard General Relativity (GR).

Varying the action (\ref{1}) with respect to the metric $g^{\mu\nu}$
yields the field equations
\begin{equation}\label{2}
\begin{array}{rcl}
&
&F_{1}(R)R_{\mu\nu}-\frac{1}{2}f_{1}(R)g_{\mu\nu}+(g_{\mu\nu}\square-\triangledown_{\mu}\triangledown_{\nu})F_{1}(R)
=-2G(L_{m})F_{2}(R)R_{\mu\nu}\\ & &
-2(g_{\mu\nu}\square-\triangledown_{\mu}\triangledown_{\nu})
G(L_{m})F_{2}(R) -f_{2}(R)[K(L_{m})L_{m}-G(L_{m})]g_{\mu\nu}
\\ & & +f_{2}(R)K(L_{m})T_{\mu\nu},
\end{array}
\end{equation}
where $\square=g^{\mu\nu}\triangledown_{\mu}\triangledown_{\nu}$,
$F_{i}(R) = df_{i}(R)/dR$ ($i = 1, 2$) and $K (L_{m}) = dG(L_{m})
/dL_{m}$. The energy-momentum tensor of matter is defined as:
\begin{equation}\label{3}
T_{\mu\nu}=-\frac{2}{\sqrt{-g}}\frac{\delta(\sqrt{-g}L_{m})}{\delta
g^{\mu\nu}}.
\end{equation}

Assuming that the Lagrangian density of matter $L_{m}$ only depends
on the metric tensor components and not on its derivatives, we
obtain
\begin{equation}\label{4}
T_{\mu\nu}=L_{m}g_{\mu\nu}-2\frac{\partial L_{m}}{\partial
g_{\mu\nu}}.
\end{equation}

The trace of the field equations (\ref{2}) reads
\begin{equation}\label{5}
\begin{array}{rcl}
& &
3\square[F_{1}(R)+2G(L_{m})F_{2}(R)]+[F_{1}(R)+2G(L_{m})F_{2}(R)]R\\
& & -2f_{1}(R)+4f_{2}(R)[K(L_{m})L_{m}-G(L_{m})] =K(L_{m})f_{2}(R)T,
\end{array}
\end{equation}
where $T=T^{\mu}_{\mu}$.

By taking the covariant divergence of Eq.(\ref{2}) and using the
mathematical identity
$\triangledown^{\mu}[f'(R)R_{\mu\nu}-\frac{1}{2}f(R)g_{\mu\nu}+(g_{\mu\nu}\square-\triangledown_{\mu}\triangledown_{\nu})f(R)]\equiv0$
\cite{29}, here $f'(R) = df/dR$, we have
\begin{equation}\label{6}
\triangledown^{\mu}T_{\mu\nu}=2\triangledown^{\mu}\ln[f_{2}(R)K(L_{m})]\frac{\partial
L_{m}}{\partial g^{\mu\nu}},
\end{equation}
from which we see that the conservation of the energy-momentum
tensor of matter is violated  due to the coupling between matter and
geometry. However, once the $L_{m}$ is given, by choosing
appropriate forms of $G(L_{m})$ and $f_{2}(R)$, one can construct,
at least in principle, conservative model with arbitrary
matter-geometry coupling.

In order to keep the energy density is positive and cannot flow
faster than light, the generalized energy conditions, namely, the
strong energy condition (SEC), the null energy condition (NEC), the
weak energy condition (WEC) and the dominant energy condition (DEC),
should be taken into consideration, which forms can be derived as
follows (see Ref.\cite{30} for more details):
\begin{equation}\label{SEC20}
\begin{array}{rcl}& &
\rho+3p-\frac{1}{f_{2}G'}[f_{1}-(f'_{1}+2Gf'_{2})R]+3\frac{f''_{1}}{f_{2}G'}(H\dot{R}+\ddot{R})\\
& & +3\frac{f'''_{1}}{f_{2}G'}\dot{R}^{2}
+6\frac{1}{f_{2}G'}(G''\dot{L_{m}}^{2}f'_{2}+\ddot{L_{m}}G'f'_{2}+2f''_{2}\dot{R}G'\dot{L_{m}}\\
& & +f'''_{2}\dot{R}^{2}G+f''_{2}\ddot{R}G)
+6\frac{H}{f_{2}G'}(G'\dot{L_{m}}f'_{2}+f''_{2}\dot{R}G)
\\ & & +\frac{2}{G'}(G'L_{m}-G)\geq0,~~~~~(SEC)
\end{array}
\end{equation}

\begin{equation}\label{NEC22}
\begin{array}{rcl}
& &
\rho+p+(H\dot{R}+\ddot{R})\frac{f''_{1}}{f_{2}G'}+\frac{f'''_{1}}{f_{2}G'}\dot{R}^{2}+\frac{2}{f_{2}G'}
(G''\dot{L_{m}}^{2}f'_{2}+\\ & & \ddot{L_{m}}G'f'_{2}+
2f''_{2}\dot{R}G'\dot{L_{m}}+f'''_{2}\dot{R}^{2}G+f''_{2}\ddot{R}G)-\\
& &
\frac{2H}{f_{2}G'}(G'\dot{L_{m}}f'_{2}+f''_{2}\dot{R}G)\geq0,~~~~~(NEC)
\end{array}
\end{equation}

\begin{equation}\label{DEC25}
\begin{array}{rcl}& &
\rho-p+\frac{1}{f_{2}G'}[f_{1}-(f'_{1}+2Gf'_{2})R]-(5H\dot{R}+\ddot{R})\frac{f''_{1}}{f_{2}G'}-\\
& & \frac{f'''_{1}}{f_{2}G'}\dot{R^{2}} -\frac{2}{f_{2}G'}
(G''\dot{L_{m}}^{2}f'_{2}+\ddot{L_{m}}G'f'_{2}+2f''_{2}\dot{R}G'\dot{L_{m}}+
\\ & &
f'''_{2}\dot{R}^{2}G+f''_{2}\ddot{R}G)-\frac{10H}{f_{2}G'}
(G'\dot{L_{m}}f'_{2}+f''_{2}\dot{R}G)-\frac{2}{G'}\\ & &
(G'L_{m}-G)\geq0,~~~~~(DEC)
\end{array}
\end{equation}

\begin{equation}\label{WEC26}
\begin{array}{rcl}& &
\rho+\frac{1}{2f_{2}G'}[f_{1}-(f'_{1}+2Gf'_{2})R]-3H\dot{R}\frac{f''_{1}}{f_{2}G'}-6H\frac{1}{f_{2}G'}
\\ & &
(G'\dot{L_{m}}f'_{2}+f''_{2}\dot{R}G)-\frac{1}{G'}(G'L_{m}-G)\geq0,~~~~~(WEC)
\end{array}
\end{equation}
where the dot denotes differentiation with respect to cosmic time.
Moreover, by using condition $f_{1}''(R)+2G(L_{m})f_{2}''(R)\geq0$,
one can test the Dolgov-Kawasaki instabilities for this class of
models.

When $G(L_{m})=L_{m}$ and rescales the function $f_{2}(R)$ as
$1+\lambda f_{2}(R)$, the action (\ref{1}) and the field equations
(\ref{2}) can be changed into
\begin{equation}\label{7}
S=\int\{\frac{1}{2}f_{1}(R)+[1+\lambda
f_{2}(R)]L_{m}\}\sqrt{-g}d^{4}x,
\end{equation}
\begin{equation}\label{8}
\begin{array}{rcl}
&
&F_{1}(R)R_{\mu\nu}-\frac{1}{2}f_{1}(R)g_{\mu\nu}+(g_{\mu\nu}\square-\triangledown_{\mu}\triangledown_{\nu})F_{1}(R)=-2\lambda
F_{2}(R)L_{m}R_{\mu\nu}\\ & &
+2\lambda(\triangledown_{\mu}\triangledown_{\nu}-g_{\mu\nu}\square)L_{m}F_{2}(R)+[1+\lambda
f_{2}(R)]T_{\mu\nu}.
\end{array}
\end{equation}
Above expressions are just the action and the field equations in
$f(R)$ gravity with non-minimal coupling between geometry and
matter. Moreover, by means of the generalized Bianchi identities
$\triangledown^{\mu}G_{\mu\nu}=0$ (here, $G_{\mu\nu}$ is the
Einstein tensor), Eq.(\ref{6}) can be given as:
\begin{equation}\label{9}
\triangledown^{\mu}T_{\mu\nu}=\frac{\lambda F_{2}}{1+\lambda
f_{2}}[g_{\mu\nu}L_{m}-T_{\mu\nu}]\triangledown^{\mu}R.
\end{equation}
It follows that the non-minimal coupling term results in a
non-trivial exchange of energy and momentum between geometry and
matter\cite{31,32}. Note that according to Eq.(\ref{9}), the
conservation of the energy-momentum tensor can be verified if
$f_{2}(R)$ is a constant or the Lagrangian density of matter is not
an explicit function of the metric.

\section{The conditions
for late-time cosmic accelerated expansion in $f(R)$ gravity with
non-minimal coupling}

~~~~In the following, we focus on the conditions for late-time
cosmic accelerated expansion in $f(R)$ gravity with non-minimal
coupling between geometry and matter. In this section, the form of
the action is taken to be Eq.(\ref{7}) and, for simplicity, we
consider $L_{m}$ is opposite to the energy density of perfect
fluid\cite{32}, i.e.,
\begin{equation}\label{33}
L_{m}=-\rho=-\rho_{0}a^{-3(1+w)},
\end{equation}
where $w$ is the equation of state of perfect fluid and is assumed
to be a constant. The energy-momentum tensor is taken as:
\begin{equation}\label{34}
T_{\mu\nu}=(\rho+p)U_{\mu}U_{\nu}+pg_{\mu\nu},
\end{equation}
where $\rho$ and $p$ denote the energy density and the pressure
respectively. The form of the FRW metric is chosen as
\begin{equation}\label{g1}
ds^{2}=-dt^{2}+a^{2}(t)dX^{2}_{3},
\end{equation}
where $a(t)$ is the scale factor and $dX^{2}_{3}$ contains the
spacial part of the metric. Using this metric, we can obtain
$R=6(2H^{2}+\dot{H})$, where $H=\dot{a}(t)/a(t)$ is the Hubble
expansion parameter.

It is known that under the conditions either power-law expansion or
the equation of state of matter less than $-1/3$, late-time cosmic
accelerated expansion occurs. To exemplify how to use these
conditions to realize the phase of accelerating expansion in $f(R)$
gravity with non-minimal coupling, Firstly, we concentrate on two
simple classes of models.

\paragraph*{~~~~(1)} Let
\begin{equation}\label{35}
f_{1}(R)=R,  ~~~~~~~~~~~~~~~~f_{2}(R)=-AR^{-n}+BR^{2},
\end{equation}
where $A$ and $B$ are arbitrary constants. Then, Eq.(\ref{8})
becomes into
\begin{equation}\label{36}
\begin{array}{rcl}
&
&3H^{2} = -\rho_{0}a^{-3(1+w)}[1+6\lambda(H^{2}+\dot{H})(24BH^{2}+12B\dot{H}+\\
& &
An(12H^{2}+6\dot{H})^{-1-n})+\lambda(36B(2H^{2}+\dot{H})^{2}-A(12H^{2}+6\dot{H})^{-n})].
\end{array}
\end{equation}
We assume the solution of Eq.(\ref{36}) is $a = a_{0}t^{p}$ and then
have $H = \frac{p}{t}$, $\dot{H} = -\frac{p}{t^{2}}$. Substituting
these relations into Eq.(\ref{36}), we find there are three kinds of
possible relationships among $p$, $w$ and $n$, namely, $p =
\frac{2(n+1)}{3(1+w)}$, $p = \frac{2}{3(1+w)}$ and $p =
\frac{-2}{3(1+w)}$. Under the condition of power-law expansion
(i.e., $p>1$), the corresponding regions of $w$ are $w <
\frac{2(n+1)}{3}-1$ for $p = \frac{2(n+1)}{3(1+w)}$, $w < -1/3$ for
$p = \frac{2}{3(1+w)}$ and $w < -5/3$ for $p = \frac{-2}{3(1+w)}$,
respectively. Furthermore, by considering the equation of state of
matter less than $-1/3$ (i.e., $w<-1/3$), we can obtain that when $p
= \frac{2(n+1)}{3(1+w)}$, the range of parameter $n$ is $n \leq 0$
and $n \neq -1$. It is easy to see that there is no constraint on
$n$ when cases $p = \frac{2}{3(1+w)}$ and $p = \frac{-2}{3(1+w)}$.
For the case $p = \frac{2(n+1)}{3(1+w)}$, the effective quintessence
regime ($-1 < w < -1/3$) emerges when $-1 < n \leq 0$ and the
effective phantom regime ($w < -1$) emerges when $n < -1$. The
candidate for late-time cosmic accelerated expansion can be either
the effective quintessence or the effective phantom, when $p =
\frac{2}{3(1+w)}$.

\paragraph*{~~~~(2)}
Another choice for functions $f_{1}(R)$ and $f_{2}(R)$ are
\begin{equation}\label{37}
f_{1}(R)=R,~~~~~~~~~ f_{2}(R)=\frac{c_{1}R^{n}}{c_{2}R^{n}+1},
\end{equation}
where $c_{1}$ and $c_{2}$ are constants. Then the FRW equation is
changed into
\begin{equation}\label{38}
3H^{2} =
-\rho_{0}a^{-3(1+w)}[1+\frac{6^{n}c_{1}n\lambda(H^{2}+\dot{H})(2H^{2}+\dot{H})^{-1+n}}{[1+6^{n}c_{2}(2H^{2}+\dot{H})^{n}]^{2}}+
\frac{6^{n}c_{1}\lambda(2H^{2}+\dot{H})^{n}}{1+6^{n}c_{2}(2H^{2}+\dot{H})^{n}}].
\end{equation}

By calculations and analysis, the relationship among $p$, $w$ and
$n$, condition and candidate for late-time cosmic accelerated
expansion are shown in Table \ref{T1}.
\begin{table}[ht]
\begin{center}
\begin{tabular}{|l|c|c|c|}
\hline Relationship & Condition  &
\multicolumn{2}{c|}{Candidate} \\
\cline{3-4}
 && \small{The effective quintessence} &\small{The effective
phantom} \\ \hline $p=\frac{2(1-n)}{3(1+w)}$ & $n\geq0$ and $n\neq1$
& $0\leq n <1$ & $n>1$ \\ \hline $p=\frac{2(1-2n)}{3(1+w)}$ &
$n\geq0$ and $n\neq1/2$ & $0\leq n <-1/2$ & $n>1/2$ \\ \hline
$p=\frac{2}{3(1+w)}$ & $w<-1/3$ &  All $n $&  All $n$ \\
\hline
\end{tabular}
\caption{\small{The relationship among $p$, $w$ and $n$, condition
and candidate for late-time cosmic accelerated expansion in case
$f_{1}(R)=R$, $f_{2}(R)=\frac{c_{1}R^{n}}{c_{2}R^{n}+1}$.}}
\label{T1}
\end{center}
\end{table}

It is worth stressing that the de Sitter stage is impossible in both
models because there is a scale factor $a$ in the FRW equation. If
the function $f_{2}(R)$ vanishes, the standard FRW equation would be
reproduced. Above forms of $f_{2}(R)$ have been discussed in
Refs.\cite{21,33}. Next, we focus on other two complicated models.

\paragraph*{~~~~(3)}
Following Ref.[21], let us take the following explicit choice for
functions $f_{1}(R)$ and $f_{2}(R)$ as:
\begin{equation}\label{39}
f_{1}(R)=R-AR^{-n}+BR^{2},~~~~~~~~~ f_{2}(R)=-AR^{-n}+BR^{2},
\end{equation}
where $A$ and $B$ are arbitrary constants. Then the equation
(\ref{8}) can be expressed as:
\begin{equation}\label{40}
\begin{array}{rcl}
& & \frac{1}{2}
\{12H^{2}+6\dot{H}+36B(2H^{2}+\dot{H})^{2}-A(12H^{2}+6\dot{H})^{-n}-6(H^{2}+\dot{H})[1+24BH^{2}+
\\ & &
12B\dot{H}+An(12H^{2}+6\dot{H})^{-1-n}]\}=-\rho_{0}a^{-3(1+w)} \{
1+6\lambda(H^{2}+\dot{H})[24BH^{2}+12B\dot{H}+
\\ & &
An(12H^{2}+6\dot{H})^{-1-n}]+\lambda[36B(2H^{2}+\dot{H})^{2}-A(12H^{2}+6\dot{H})^{-n}]\}.
\end{array}
\end{equation}

By the same method as above, we find there are four kinds of
possible relationships among $p$, $w$ and $n$, i.e.,
$p=\frac{4}{3(1+w)}$, $p=-\frac{2n}{4+3w}$, $p=\frac{2}{3(1+w)}$ and
$p=-2n$. Under conditions of power-law expansion and the equation of
state of matter less than $-1/3$, above results are turned into
$w<1/3$ for $p=\frac{4}{3(1+w)}$, $n\geq-3/2$ for
$p=-\frac{2n}{4+3w}$, $w<-1/3$ for $p=\frac{2}{3(1+w)}$ and $n<-1/2$
for $p=-2n$, respectively. It is clear that the candidate for
late-time cosmic accelerated expansion is among dust, the effective
quintessence and the effective phantom when $p=\frac{4}{3(1+w)}$.
But except dust when $p=\frac{2}{3(1+w)}$. Note that there is no
constraint on $n$ in both cases. For the case $p=-\frac{2n}{4+3w}$,
either the effective quintessence regime emerges when $-3/2\leq
n<-1/2$ or the effective phantom regime emerges when $n\geq-1/2$ and
$n\neq0$. Obviously, the late-time cosmic accelerated expansion is
independent of matter when $p=-2n$.

\paragraph*{~~~~(4)}
Another choice for functions $f_{1}(R)$ and $f_{2}(R)$ are
\begin{equation}\label{41}
f_{1}(R)=R+\frac{c_{1}R^{n}}{c_{2}R^{n}+1},~~~~~~~~~
f_{2}(R)=\frac{c_{1}R^{n}}{c_{2}R^{n}+1},
\end{equation}
where $c_{1}$ and $c_{2}$ are constants. In this case, the FRW
equation can be given as:
\begin{equation}\label{42}
\begin{array}{rcl}
& &
\frac{1}{2}[12H^{2}+6\dot{H}+\frac{6^{n}c_{1}(2H^{2}+\dot{H})^{n}}{1+6^{n}c_{2}(2H^{2}+\dot{H})^{n}}-6(H^{2}+\dot{H})
(1+\frac{6^{-1+n}c_{1}n(2H^{2}+\dot{H})^{-1+n}}{[1+6^{n}c_{2}(2H^{2}+\dot{H})^{n}]^{2}})]=
\\\\ & & -\rho_{0}a^{-3(1+w)}[1+\frac{6^{n}c_{1}n\lambda(H^{2}+\dot{H})(2H^{2}+\dot{H})^{-1+n}}{[1+6^{n}c_{2}(2H^{2}+\dot{H})^{n}]^{2}}
+\frac{6^{n}c_{1}\lambda(2H^{2}+\dot{H})^{n}}{1+6^{n}c_{2}(2H^{2}+\dot{H})^{n}}].
\end{array}
\end{equation}

The corresponding relationship among $p$, $w$ and $n$, condition and
candidate for late-time cosmic accelerated expansion are shown in
Table \ref{T2}.

\begin{table}[ht]
\begin{center}
\begin{tabular}{|l|c|c|c|}
\hline Relationship & Condition &
\multicolumn{2}{c|}{Candidate} \\
\cline{3-4}
 &  & \small{The effective quintessence} &\small{The effective
phantom} \\ \hline $p=\frac{2}{3(1+w)}$& $w<-1/3$ & All $n$ & All
$n$ \\ \hline $p=\frac{4n+2}{3(1+w)}$ & $n\leq0$ and $n\neq-1/2$ &
$-1/2<n\leq0$ & $n<-1/2$ \\ \hline $p=\frac{2(n+1)}{3(1+w)}$&
$n\leq0$ and $n\neq-1$ & $-1<n\leq0$& $n<-1$
 \\ \hline
 $p=\frac{2n}{3(1+w)}$ & $n\leq1$ and $n\neq0$ & $0<n\leq1$ & $n<0$
 \\ \hline
$p=\frac{4n}{3(1+w)}$& $n\leq1/2$ and $n\neq0$ & $0<n\leq1/2$ &
$n<0$ \\ \hline
\end{tabular}
\caption{\small{The relationship among $p$, $w$ and $n$, condition
and candidate for late-time cosmic accelerated expansion in case
$f_{1}(R)=R+\frac{c_{1}R^{n}}{c_{2}R^{n}+1}$,
$f_{2}(R)=\frac{c_{1}R^{n}}{c_{2}R^{n}+1}$.}} \label{T2}
\end{center}
\end{table}

Obviously, the de Sitter stage is also impossible in both models and
the reason is as the same as the ones in above two simple cases. If
the function $f_{2}(R)$ vanishes, the modified gravity with
non-coupling can be reproduced. Above forms of $f_{1}(R)$ have been
discussed in Refs.\cite{21,34}.

From the above discussions, it is easy to see that the results in
complicated models are more interesting than the simple ones.

For the four models mentioned above, the transition from matter-
dominated phase to the acceleration phase ㄗdiscussions without
non-minimal coupling have been made in Ref.\cite{35}ㄘcould be
realized as follows. Since the Hubble parameter can be expressed as
$H=p/t$, Ricci scalar $R$ turns into $R=6p(2p-1)/t^2$. If $0<p<1$,
the early universe is in deceleration phase, which corresponds to
matter-dominated phase with $p=2/3$, and if $p>1$, the late universe
is in acceleration phase.

\section{The instability of $f(R)$ gravity with
non-minimal coupling}

~~~~A viable modified gravity model must pass Newton law, solar
system test and instability conditions\cite{21,36}. There are in
principle several kinds of instabilities to consider\cite{37}.
Dolgov每Kawasaki instability\cite{38} is one of them. Below, we will
focus on this instability. According to Ref.\cite{30}, the
Dolgov每Kawasaki criterion in $f(R)$ gravity with non-minimal
coupling between matter and geometry is
\begin{equation}\label{43}
f_{1}^{\prime\prime}(R)+2\lambda L_{m}f_{2}^{\prime\prime}(R)\geq0
\end{equation}
where $\lambda$ is a constant and $L_{m}$ is the Lagrangian density
of matter. For simplicity assuming $A$, $B$, $C_{1}$, $C_{2}$ are
positive constants, the Dolgov每Kawasaki criterions for above four
discussed models are as follows:
\begin{equation}\label{44}
n\leq0 ~~or~~ n\leq-1, ~~~~for~~ model~~ 1~~ and ~~model~~ 3,
\end{equation}

\begin{equation}\label{45}
n\leq\frac{1+C_{2}R^{n}}{1-C_{2}R^{n}}, ~~~~for~~ model~~ 2 ~~and
~~model~~ 4
\end{equation}

By means of analysis of model 1, model 3 and Eq.(\ref{44}), we find
that they could be realistic candidates for late-time cosmic
accelerated expansion without Dolgov每Kawasaki instability. By
taking Table \ref{T1}, Table \ref{T2} and Eq.(\ref{45}) into
consideration, model 2 and model 4 would be realistic candidates if
$1+C_{2}R^{n}/1-C_{2}R^{n}>0$ and $1+C_{2}R^{n}/1-C_{2}R^{n}\leq1$
respectively. Otherwise there is no interesting in model 2.
Furthermore, the Dolgov每Kawasaki instability will be emerged in
model 4, if $1+C_{2}R^{n}/1-C_{2}R^{n}<1/2$ (i.e. $1/2\leq n\leq1$)
for case $p=2n/(1+w)$, $0\leq n\leq1$ for case $p=2n/(1+w)$ and
$0\leq n\leq1/2$ for case $p=4n/(1+w)$, respectively.

\section{Summary}

~~Up to now, we have discussed the conditions for late-time cosmic
accelerated expansion and the Dolgov每Kawasaki instability in $f(R)$
gravity with non-minimal coupling between geometry and matter. For
simplicity, we chose the form of the Lagrangian density of matter as
opposite to the energy density of perfect fluid. The relationship
among $p$, $w$ and $n$ has been given in each class of models. By
using the conditions of power-law accelerated expansion, the
equation of state of matter less than $-1/3$ and the
Dolgov每Kawasaki criterion, the range of the parameter $n$ is
concretely constrained. Either the effective quintessence regime or
the effective phantom regime would emerge by choosing $n$ properly.
It is easy to see that the results in complicated models are more
interesting than the simple ones. It is demonstrated that the de
Sitter stage would not realize in all considering models because
there is a scale factor $a$ in FRW equation. Essentially, this is
due to the special choice of the Lagrangian density of matter. Other
forms of the Lagrangian density of matter could be considered in the
similar fashion of non-minimal gravitational coupling.

\end{document}